\documentclass[12pt]{article}
\begin{document}
\title{{\bf DEFINING ENTROPY BOUNDS}
\thanks{Alberta-Thy-09-00, hep-th/0007238}}
\author{
Don N. Page
\thanks{Internet address:
don@phys.ualberta.ca}
\\
Theoretical Physics Institute\\
Department of Physics, University of Alberta\\
Room 238 CEB, 11322 -- 89 Avenue\\
Edmonton, Alberta, Canada T6G 2G7
}
\date{(2008 Sept. 19)}
\maketitle
\large

\begin{abstract}
\baselineskip 16 pt

	Bekenstein's conjectured entropy bound for a system of linear
size $R$ and energy $E$, $S \leq 2 \pi E R$, has counterexamples for
many of the ways in which the ``system,'' $R$, $E$, and $S$ may be
defined. Here new ways are proposed to define these quantities for
arbitrary nongravitational quantum field theories in flat spacetime,
such as defining $R$ as the smallest radius outside of which only vacuum
expectation values occur. Difficulties of extending these definitions to
gravitational quantum and semiclassical theories are noted.

\end{abstract}
\normalsize
\baselineskip 16 pt
\newpage

\section{Introduction}

	Bekenstein has conjectured \cite{Bek1} that the entropy $S$ of a
system confined to radius $R$ or less and energy $E$ or less would obey
the inequality (using units $\hbar = c = k_{\mathrm Boltzmann} = 1$)
 \begin{equation}
 S \leq 2 \pi E R.
 \label{eq:1}
 \end{equation}
He and colleagues have supported this conjecture with many arguments 
and examples
[1-18].
However, many counterarguments and counterexamples have also been noted
[19-34].
Whether the conjectured bound (\ref{eq:1}) holds or not depends on what
systems are considered and how $R$, $E$, and $S$ are defined.

	Perhaps the simplest procedure \cite{SB1,BS} would be to just
consider quantum fields inside some bounded region within a sphere of
radius $R$ and put boundary conditions on the fields at the boundary of
the region.  However, this procedure leads to a large number of
counterexamples to Bekenstein's conjectured bound.  For example
\cite{Page1}, the Casimir effect can make $E<0$ for certain states of
quantum fields confined within a certain regions of radius $\leq R$,
violating the bound. If states with $E<0$ are excluded by definition,
one can still consider a mixed state with arbitrarily small positive $E$
that violates the bound. Even if $E$ is redefined to be the nonnegative
energy excess over that of the ground state \cite{SB1,BS}, one can
violate the bound by a mixed state that is almost entirely the ground
state and a tiny incoherent mixture of excited states, at least if the
entropy is defined to be $S = - tr\rho\ln\rho$ \cite{Deu}. If $S$ is
instead defined to be $S = \ln n$ for a mixture of $n$ orthogonal pure
states (which would agree with $S = - tr\rho\ln\rho$ if the mixture had
equal probabilities $1/n$ for each of those $n$ pure states), then one
can violate the bound by an equal mixture of the ground state and the
first excited state of certain interacting fields with certain boundary
conditions that have the two lowest states nearly degenerate in energy
(separated by exponentially small tunneling effects)
\cite{Page1,Page2,Page3}. If interacting fields are excluded from the
definition of allowable systems, one can get a violation by considering
a sufficiently large number $N$ of identical free fields, giving $n=N$
degenerate first excited states of finite energy but sufficiently large
entropy $S = \ln n = \ln N$ to violate (\ref{eq:1}) \cite{Page1,UW1}.
And even for a single free electromagnetic field, $S = \ln n$ can exceed
$2 \pi E R$ by an arbitrarily large factor by using boundary conditions
corresponding to an arbitrarily large number of parallel perfectly
conducting plates within the region of radius $R$ \cite{Page2}, or by
using boundary conditions corresponding to an arbitrarily long coaxial
cable loop coiled up within the region \cite{Page3}.

	However, other than in his papers with Schiffer \cite{SB1,BS},
Bekenstein has generally advocating taking $E$ to be the total energy of
a complete system \cite{Bek1,Bek3,Bek4,Bek6,Bek10,Bek13}.  This would
disallow using just the energy of fields within a bounded region with
boundary conditions, since that would ignore the energy of whatever it
is that is providing the boundary conditions.  Therefore, all of the
counterexamples mentioned above would be excluded by this restriction. 
However, then the problem is to define what one means by the radius $R$
of the system.  In the weakly gravitating case (essentially quantum
fields in flat Minkowski spacetime) that we shall focus on here,
Bekenstein takes $R$ to mean the radius of a sphere which circumscribes
the system, which leaves the problem of what it means for a sphere to
circumscribe the complete system.

	In quantum field theory in Minkowski spacetime, the complete
system is the quantum state of the fields.  Since the quantum fields
extend all the way out to radial infinity, a sphere circumscribing the
entire system would have to be at $R=\infty$, which makes the Bekenstein
bound true (at least for states of positive energy and finite entropy)
but trivial.  To get a nontrivial bound, one needs to suppose that a
sphere of finite $R$ can circumscribe the system.  For example, one
might try to say that the sphere encloses all of the excitations of the
fields from the vacuum.  However, it is also hard to get this to occur
for a finite $R$.  For example, the wavefunction for any single particle
state that is a superposition of energy eigenstates of bounded energy
will not vanish outside any finite radius $R$, since a one-particle
wavefunction that does vanish outside a finite region must be a
superposition of arbitrarily large momentum components, which will have
unbounded energy.  Even if one looks at a composite system, such as a
hydrogen atom, and ignores the fact that its center of mass will have
amplitudes to be outside any finite sphere if it is made of purely
bounded energy components, the wavefunction for the relative position of
the electron and proton does not drop identically to zero outside any
finite separation distance for states that are superpositions of energy
eigenstates of bounded energy.  In particular, even if one fixed the
center of mass of a hydrogen atom in its ground state and ignored the
infinite energy from the resulting infinite uncertainty of the center of
mass momentum, the density matrix for the electron position would decay
only exponentially with distance from the center of mass and never go to
zero outside any sphere of finite radius $R$.

	Therefore, it is problematic to define the radius $R$ of a
sphere circumscribing a complete system in any quantum field theory. 
This issue has not been addressed by Bekenstein and his collaborators,
but without such a definition, there is no nontrivial formulation of the
conjectured bound (\ref{eq:1}) for complete systems, but only its
trivial truth for any complete system with positive energy and finite
entropy that can only be circumscribed by the sphere enclosing all of
space, $R=\infty$.

	Here new ways are proposed to define systems and their radii
$R$, energies $E$, and entropies $S$, so that for each, there is a bound
on $S$ for a given system as a function of finite $R$ and $E$. These
bounds will not have the form of Bekenstein's conjectured inequality
(\ref{eq:1}), though in some cases they may obey that inequality.

\section{Vacuum-Outside-R States}

	The main new element of the present paper is a proposal is to
define a system of radius $R$ (in flat spacetime for the present) not by
imposing boundary conditions on the field itself, but by imposing
conditions on the quantum state of the field so that outside a closed
ball of radius $R$ the quantum state is indistinguishable from the
vacuum at some time. Such a state will be called a vacuum-outside-$R$
state. (For simplicity, set this time to be $t=0$, and take the closed
ball, say $B$, to be the region $r\leq R$ on the $t=0$ hypersurface,
where $r$ is the standard radial polar coordinate giving the proper
distance from the coordinate origin on that hypersurface.) In other
words, a vacuum-outside-$R$ state of the system, say as expressed by its
density matrix $\rho$, is such that the expectation value of any
operator $O$ which is completely confined to the region $r>R$ when
written in terms of field and conjugate operators at $t=0$, is precisely
the same as the expectation value of the same operator in the vacuum
state $|0\!><\!0|$,
 \begin{equation}
 tr(O\rho) = <\!0|O|0\!>.
 \label{eq:2}
 \end{equation}
In particular, all the $n$-point functions for the field and for its
conjugate momentum in the state $\rho$ are the same as in the vacuum
state, if all of the $n$ points are outside the ball of radius $R$ and
on the hypersurface $t=0$. Of course, the $n$-point functions need not
be the same as their vacuum values if some or all of the points are
inside the ball.

	If operators confined to the three-dimensional region $r>R$ and
$t=0$ (say $C$, to give a name to this achronal spacelike surface, the
$t=0$ hypersurface with the central closed ball $B$, $r\leq R$,
excluded) have the same expectation value in the vacuum-outside-$R$
state as in the vacuum state, the same will be true in any quantum field
theory that I shall call ``strongly causal'' for all operators confined
to the Cauchy development or domain of dependence \cite{Haw} of $C$, the
larger four-dimensional region $r>R+|t|$ (say $D$) that is the set of
all points in the Minkowski spacetime such that every inextendible
(endless) causal, or non-spacelike (everywhere timelike or lightlike),
curve through such a point intersects the partial Cauchy surface $C$.
Just as solutions of hyperbolic wave equations in $D$ are determined by
the data on $C$, so the part of the quantum state of a strongly causal
field in $D$, as represented by the expectation values of operators
confined to $D$, is determined by the part of the quantum state in $C$,
as represented by the expectation values of operators confined to $C$.
(For some interacting quantum field theories, the expectation values of
operators confined to the three-dimensional spacelike surface $C$ may be
too ill-defined for these theories to be ``strongly causal'' in my
sense, but a wider class of these theories may be ``weakly causal'' in
the sense that sufficiently many operators smeared over, but confined
to, an arbitrarily thin-in-time four-dimensional slab, say $E$,
containing $C$ within $D$, have well-defined expectation values that
determine the expectation values of all operators smeared over, but
confined to, any part of $D$.)

	Henceforth I shall restrict attention to strongly causal and
weakly causal quantum field theories, calling them simply causal quantum
field theories for short. I shall also assume, until discussing
gravitational theories later, that any quantum field theory under
consideration is a nongravitational Lorentz-invariant quantum field
theory in Minkowski spacetime, and that it has a unique pure state of
lowest Minkowski energy $E = 0$ (the expectation value of the
Hamiltonian $H$ that generates translations in the time coordinate $t$
in some Lorentz frame, with the arbitrary constant in the Hamiltonian
being adjusted to give the lowest energy state zero energy).

	Therefore, for such a causal nongravitational quantum field
theory in Minkowski spacetime, I shall propose that the radius $R$ be
defined so that all of the operators constructed from field and
conjugate momentum operators smeared over regions confined to the region
$D$, $r>R+|t|$ in some Lorentz frame, have in the particular quantum
state being considered (a vacuum-outside-$R$ state) the same expectation
values that they have in the vacuum state for that quantum field theory.
The energy $E$ of the state $\rho$ can then be simply defined to be the
expectation value,
 \begin{equation}
 E \equiv tr(H\rho),
 \label{eq:3}
 \end{equation}
of the Hamiltonian $H$ that generates time translations in the same
Lorentz frame. Because the energy $E$ has been defined to have the
minimum value of zero for the unique pure vacuum state, there is no
problem here with negative Casimir energies. In other words, the energy
is that of the complete system over all of Minkowski spacetime.

	Obviously we would also like a definition of
the entropy $S$ that has a minimum value of zero,
which it should attain for the pure vacuum state.
One simple definition is the von Neumann entropy,
 \begin{equation}
 S = S_{\rm vN} \equiv - tr\rho\ln\rho,
 \label{eq:4}
 \end{equation}
using the density matrix $\rho$ for the full state
of the quantum field, over the entire Minkowski spacetime.

\section{Entropy Bounds for Vacuum-Outside-R States}

	Now we may conjecture that for any vacuum-outside-$R$ state of
any particular causal nongravitational quantum field theory
in Minkowski spacetime, one which has the vacuum expectation
values in the region $D$, $r>R+|t|$ (the region causally
disconnected from the ball $r \leq R$ at $t = 0$),
the von Neumann entropy is bounded above by some function
$\sigma_{\rm vN}$ (depending on the quantum field theory in question)
of the radius $R$ and energy $E$:
 \begin{equation}
 S_{\rm vN} \leq \sigma_{\rm vN}(R,E).
 \label{eq:5}
 \end{equation}
Define this function $\sigma_{\rm vN}(R,E)$ to be the
least upper bound on the von Neumann entropy
of any state which is vacuum outside the radius $R$
and which has energy $E$.
 
	In the case of a scale-invariant quantum field,
such as a free massless field, or say a massless scalar
field $\phi$ with a $\lambda \phi^4$ self-coupling potential,
the least upper bound function $\sigma_{\rm vN}(R,E)$
will actually be a function of the single dimensionless
variable
 \begin{equation}
 x \equiv 2 \pi R E,
 \label{eq:6}
 \end{equation}
say
 \begin{equation}
 \sigma_{\rm vN}(R,E) = \sigma_{\rm N}(x).
 \label{eq:7}
 \end{equation}

	Bekenstein's conjectured entropy bound (\ref{eq:1}),
if $R$, $E$, and $S$ were defined as done herein,
would be $\sigma_{\rm vN}(R,E) \leq x$,
whether or not the quantum field theory
is scale invariant, or
 \begin{equation}
 B_{\rm vN}(R,E) \equiv {\sigma_{\rm vN}(R,E) \over x}
                 \equiv {\sigma_{\rm vN}(R,E) \over 2\pi R E} \leq 1.
 \label{eq:7b}
 \end{equation}
If the quantum field theory is scale invariant,
we can define
 \begin{equation}
 B_{\rm N}(x) \equiv {\sigma_{\rm N}(x) \over x},
 \label{eq:7c}
 \end{equation}
which should also be less than or equal to unity
if Bekenstein's bound applies.

	For a set of one or more free massless fields
and vacuum-outside-$R$ states with $x \gg 1$,
one would expect that the highest entropy would
be given by a mixed state that at $t=0$ is approximately
a high-temperature ($RT \gg 1$) thermal radiation state
for $r < R$, surrounded by vacuum for $r > R$.
A high-temperature thermal radiation state has an
energy density for massless fields of approximately $a_r T^4$,
and hence an entropy density $(4/3)a_r T^3$, where
 \begin{equation}
 a_r = {\pi^2\over 30}(n_b + {7\over 8}n_f)
 \label{eq:8}
 \end{equation}
is the radiation constant for $n_b$ independent
bosonic degrees of freedom for each momentum
(e.g., $n_b$ different spin or helicity states)
and for $n_f$ fermionic degrees of freedom.
Therefore, in this case with $x \gg 1$,
 \begin{equation}
 B_{\rm vN}(R,E) = {\sigma_{\rm N}(x) \over x}
 \approx \left({2^7 a_r \over 3^5 \pi^2 x} \right)^{1\over 4}
 = \left[{2^6\over 3^6 5 x}(n_b + {7\over 8}n_f)\right]^{1\over 4},
 \label{eq:9}
 \end{equation}
which is indeed less than 1, thus obeying Bekenstein's
conjectured bound, for
 \begin{equation}
 x \geq {2^7 a_r \over 3^5 \pi^2}
   = {2^6\over 3^6 5}(n_b + {7\over 8}n_f)
   = {64 n_b + 56 n_f \over 3645},
 \label{eq:10}
 \end{equation}
if $x$ is also large enough that Eq. (\ref{eq:9}) is a good
approximation. Thus one would expect that Bekenstein's conjectured
bound, using the definitions above for $R$, $E$, and $S$, holds for a
fixed set of free massless quantum fields at sufficiently large $x
\equiv 2 \pi R E$.

	On the other hand, the definitions above
for $R$, $E$, and $S$ still permit Bekenstein's conjectured
bound applied to them to be violated for sufficiently small $x$,
as we can see by the following construction:

	A way to construct vacuum-outside-$R$ states,
quantum states of a free quantum field
theory that have vacuum expectation values in the region $D$,
$r>R+|t|$, is to apply to the vacuum state unitary operators
constructed from fields and/or conjugate momenta smeared within
the region $r<R$ at $t=0$.
In particular, if $h$ is an hermitian operator
constructed from fields and/or conjugate momenta
smeared within $r<R$ at $t=0$, then $U = e^{ih}$
is such a unitary operator, and
 \begin{equation}
 |\psi\!> = U|0\!> = e^{ih}|0\!>
 \label{eq:11}
 \end{equation}
is a pure quantum state that has precisely the vacuum expectation
values in the region $D$.  This result can be seen formally from
the fact that any operator $O$ confined to the region $D$
(the four-dimensional region $r>R+|t|$)
that is causally disconnected from the ball $B$
(the three-dimensional region $r\leq R$ on the $t=0$ hypersurface)
commutes with the operators $h$ and $U$ that are confined
to that hypersurface, $[O,h] = [O,U] = 0$, so
 \begin{equation}
 tr(O\rho) = <\!\psi|O|\psi\!>
           = <\!0|U^{-1}OU|0\!> = <\!0|U^{-1}UO|0\!> = <\!0|O|0\!>.
 \label{eq:12}
 \end{equation}
 
	If $\{h_i\}$ is a set of hermitian operators
that each are confined to the ball $B$ (i.e., are constructed
from fields and momenta that are smeared only over that region),
and if $\{q_i\}$ is a set of positive numbers that sum to unity, then
 \begin{equation}
 \rho = \sum_i q_i e^{i h_i}|0\!><\!0|e^{-i h_i}
 \label{eq:13}
 \end{equation}
is a more general vacuum-outside-$R$ state,
since this density matrix gives vacuum expectation values,
$tr(O\rho) = <\!0|O|0\!>$, for any operator $O$ confined to the region
$D$ that is causally disconnected from $B$
(i.e., having no causal curves, either timelike or lightlike,
intersecting both the ball $B$ of $r\leq R$ at $t=0$
and the region $D$ with $r>R+|t|$).

	Here let us consider the simple example in which
$i$ takes only the two values 1 and 2, and $h_1=0$ and $h_2=h$.
Let $q_1=1-q$ and $q_2=q$, and let
 \begin{equation}
 e^{ih}|0\!> = U|0\!> = |\psi\!> = c|0\!>+s|1\!>
 \label{eq:14}
 \end{equation}
in terms of a decomposition of $|\psi\!>$ into the two
orthonormal states $|0\!>$ and
$|1\!>=(|\psi\!>-<\!0|\psi\!>|0\!>)/\sqrt{1-|\!\!<\!0|\psi\!>\!\!|^2}$, 
so
 \begin{equation}
 c = <\!0|\psi\!> = <\!0|U|0\!> = <\!0|e^{ih}|0\!>,
 \label{eq:15}
 \end{equation}
 \begin{equation}
 s = \sqrt{1-|\!\!<\!0|U|0\!>\!\!|^2}
   = \sqrt{1-|c|^2}.
 \label{eq:16}
 \end{equation}
Note that $|1\!><\!1|$ by itself is not generically
a vacuum-outside-$R$ state.

	Now Eq. (\ref{eq:13}) gives the density matrix as
 \begin{eqnarray}
 \rho &=& (1-q)|0\!><\!0|+q|\psi\!><\!\psi| \nonumber \\
      &=& (1-qs^2)|0\!><\!0|+qcs|0\!><\!1|
      +q\bar{c}s|1\!><\!0|+qs^2|1\!><\!1|,
 \label{eq:17}
 \end{eqnarray}
a density matrix in the two-dimensional space
of pure states spanned by the two orthonormal pure
states $|0\!>$ and $|1\!>$.  The two eigenvalues of
this density matrix are, say, $p$ and $1-p$
(since their sum is $tr\rho=1$), with product
 \begin{eqnarray}
 y \equiv p(1-p) &=& {1\over 2}\{[p+(1-p)]^2-[p^2+(1-p)^2]\}
        = {1\over 2}\{[tr(\rho)]^2-[tr(\rho^2)]\} \nonumber \\
	&=& q(1-q)(1-|\!\!<\!0|U|0\!>\!\!|^2) = q(1-q)s^2.
 \label{eq:18}
 \end{eqnarray}

	The expectation value of the energy of this
mixed state is, since I have assumed $H|0\!>=0$,
 \begin{equation}
 E = tr(H\rho) = q<\!\psi|H|\psi\!> = q<\!0|U^{-1}HU|0\!>.
 \label{eq:19}
 \end{equation}
Then
 \begin{equation}
 x \equiv 2 \pi R E = 2 \pi R q<\!0|U^{-1}HU|0\!>.
 \label{eq:20}
 \end{equation}

	The von Neumann entropy of this mixed state is
 \begin{eqnarray}
 S_{\rm vN}(y) &=& - tr(\rho\ln\rho) = -p\ln{p}-(1-p)\ln{(1-p)}
  		 \nonumber \\
               &=& {y \over {1\over 2}(1+\sqrt{1-4y})} \ln{{1\over y}}
	        + \sqrt{1-4y} \ln{1\over {1\over 2}(1+\sqrt{1-4y})}
		 \nonumber \\
	       &\approx& y[(1+y+2y^2)\ln{1\over y}
	        +(1-{1\over 2}y-{5\over 3}y^2)],	
 \label{eq:21}
 \end{eqnarray}
a monotonically increasing function of $y\equiv q(1-q)s^2 \leq 1/4$,
where the last approximate equality of Eq. (\ref{eq:21}) applies for 
$y \ll 1$.

	As $q$ and/or $h$ is reduced toward zero,
$x$, $y$, and $S$ also decrease toward zero, but whereas
$x$ and $y$ asymptotically decrease linearly with $q$,
the dominant term of
$S$ has an extra logarithmic factor that grows with
the reduction of $y$, so the ratio,
 \begin{equation}
 B \equiv {S_{\rm vN}\over x} \equiv {S_{\rm vN}\over 2\pi R E}
    \approx {y\over x}\left(\ln{1\over y} + 1 \right)
 \label{eq:22}
 \end{equation}
when $y \ll 1$,
increases without limit as $y$ is reduced toward zero.
Therefore, when $y$ is made sufficiently small
(e.g., by making $q$ sufficiently small),
Bekenstein's conjectured bound for the definition of
$R$, $E$, and $S$ used here is violated.

\section{Free Quantum Field Theory Examples}

	Let us consider a specific example for
the Hermitian operator $h$ that is constructed
from field operators confined to the ball $B$
of radius $R$
on an initial flat hypersurface of Minkowski spacetime.
Take the quantum field theory to be that of a single
massless scalar field operator $\phi$.
Consider the smeared linear Hermitian field operator
 \begin{equation}
 \chi = \int d^3 x \: [F(\mathbf{x}) \: \phi(t\!=\!0,\mathbf{x})
 	+ G(\mathbf{x}) \: \dot{\phi}(t\!=\!0,\mathbf{x})],
 \label{eq:23}
 \end{equation}
where $F(\mathbf{x})$ and $G(\mathbf{x})$
are real functions of the spatial location $\mathbf{x}$
that are zero for $|\mathbf{x}|>R$,
so that $\chi$ is made up of operators confined
to the ball $B$, $r \leq R$ at $t=0$.
Then for real parameters $\alpha$ and $\beta$, let
 \begin{equation}
 h = \alpha \chi + \beta \chi^2,
 \label{eq:24}
 \end{equation}
which is thus also an Hermitian operator confined to the ball $B$.

	Then by expanding out $\phi(\mathbf{x})$ and $H$
in terms of creation and annihilation operators,
one can show, after a certain amount of algebra
that will not be repeated here, that
 \begin{equation}
 c \equiv <\!0|\psi\!> \equiv <\!0|U|0\!> \equiv <\!0|e^{ih}|0\!>
   = (1-2i\beta X)^{-1/2}\exp{\left({-{1\over 2}\alpha^2 X
   \over 1-2i\beta X}\right)},
 \label{eq:25}
 \end{equation}
 \begin{equation}
 s^2 \equiv 1-|\!\!<\!0|U|0\!>\!\!|^2 \equiv 1-|c|^2
   = 1 - (1+4\beta^2 X^2)^{-1/2}\exp{\left({-\alpha^2 X
   \over 1+4\beta^2 X^2}\right)},
 \label{eq:25b}
 \end{equation}
and
 \begin{equation}
 <\!\psi|H|\psi\!> \equiv <\!0|U^{-1}HU|0\!>
  \equiv <\!0|e^{-ih}He^{ih}|0\!>
  = \alpha^2 Y + 4 \beta^2 X Y
  = (\alpha^2 X + 4 \beta^2 X^2) Z,
 \label{eq:26}
 \end{equation}
where
 \begin{equation}
 X \equiv <\!0|\chi^2|0\!>
  = \int d^3 x d^3 y {F(\mathbf{x}) F(\mathbf{y})
  + \mathbf{\nabla}G(\mathbf{x})\!\cdot\!\mathbf{\nabla}G(\mathbf{y})
             \over 4\pi^2 |\mathbf{x}-\mathbf{y}|^2}
 \label{eq:27}
 \end{equation}
and
 \begin{equation}
 Y \equiv X Z \equiv <\!0|\chi H \chi |0\!> =
 {1 \over 2}\int d^3 x [|F(\mathbf{x})|^2 + |\nabla G(\mathbf{x})|^2].
 \label{eq:28}
 \end{equation}

	Incidentally, I have not included
individual higher powers of $\chi$ in $h$,
because then expanding $e^{ih}$ into a power series in $\chi$
and taking the expectation values gives divergent series
when one uses the key intermediate results
 \begin{equation}
 <\!0|\chi^m|0\!> = \left\{ \begin{array}{ll} 
 	(m-1)!! \: X^{m/2} & \mbox{$m$ even} \\
	0		& \mbox{$m$ odd}
		\end{array} \right. ,
 \label{eq:29}
 \end{equation}
and
 \begin{equation}
 <\!0|\chi^m H \chi^n|0\!> = \left\{ \begin{array}{ll} 
 	m n (m+n-3)!! \: X^{(m+n)/2} Z & \mbox{$m+n$ even} \\
	0		& \mbox{$m+n$ odd}
		\end{array} \right. .
 \label{eq:30}
 \end{equation}
(The divergences arise from the rapid growth
of the double factorials with their arguments.
These double factorials arise from the counting
of the number of pairings of the creation
and annihilation operators in the powers
of the $\chi$'s and in the Hamiltonian $H$
for the massless scalar field $\phi$.)

	If one takes $F(\mathbf{x})$ and $G(\mathbf{x})$
to be spherically symmetric, say
 \begin{equation}
 F(\mathbf{x}) = R^{-2}\:f\left({|\mathbf{x}|\over R}\right)
 \equiv R^{-2} f(u)
 \label{eq:31}
 \end{equation}
and
 \begin{equation}
 G(\mathbf{x}) = R^{-1}\:g\left({|\mathbf{x}|\over R}\right)
 \equiv R^{-1} g(u)
 \label{eq:32}
 \end{equation}
with $f$ and $g$ being dimensionless functions of the
dimensionless radial variable (hereafter to be called $u$ or $v$)
that vanish when the latter variable is greater
than unity (corresponding to points outside
the sphere of radius $R$), then
 \begin{equation}
 X = \int_0^1 du \int_0^1 dv \left\{
 2uv\ln{\left|{u+v\over u-v}\right|}f(u)f(v)
 +\left[(u^2+v^2)\ln{\left|{u+v\over u-v}\right|}-2uv\right]g'(u)g'(v)
 \right\}
 \label{eq:33}
 \end{equation}
and
 \begin{equation}
 Y \equiv X Z = {2\pi\over R}\int_0^1 u^2 du [f^2(u) + g'^2(u)],
 \label{eq:34}
 \end{equation}
where the prime on the function $g$ denotes a derivative
with respect to the argument
(the dimensionless radius $u \equiv |\mathbf{x}|/R$
or $v \equiv |\mathbf{y}|/R$).

	Now, if we take a density matrix of the form (\ref{eq:17}),
let us try to maximize the product of the two nonzero eigenvalues
of the density matrix,
 \begin{equation}
 y \equiv p(1-p) = q(1-q)s^2
  = q(1-q) \left[ 1 - (1+4\beta^2 X^2)^{-1/2}
  \exp{\left({-\alpha^2 X \over 1+4\beta^2 X^2}\right)} \right],
 \label{eq:35}
 \end{equation}
and hence maximize $S_{\rm vN}(y)$ given by Eq. (\ref{eq:21}),
for fixed
 \begin{equation}
 x \equiv 2\pi R E = 2 \pi R q<\!0|U^{-1}HU|0\!>
   = 2\pi R Z q (\alpha^2 X + 4 \beta^2 X^2).
 \label{eq:36}
 \end{equation}
Note that for fixed $RZ = RY/X$, the three quantities $\alpha$, $\beta$
(the coefficients of $\chi$ and of $\chi^2$ in the hermitian operator $h
= \alpha \chi + \beta \chi^2$), and $X \equiv <\!0|\chi^2|0\!>$ enter
into this $x$ and $y$ only in the two nonnegative combinations $a \equiv
\alpha^2 X$ and $b \equiv 4 \beta^2 X^2$, and $x$ depends only on
$q(a+b)$. Then it is easy to see that for fixed $q$ and fixed $a+b$, $y$
decreases monotonically with $b$, so to maximize $y$ and $S_{\rm vN}(y)$
for fixed $x$, we should set $\beta = 0$ in order to get $b=0$, $h =
\alpha \chi$,
 \begin{equation}
 a \equiv \alpha^2 X = <\!0|h^2|0\!>,
 \label{eq:37}
 \end{equation}
 \begin{equation}
 x = 2\pi R Z q a,
 \label{eq:38}
 \end{equation}
and
 \begin{equation}
 y = q(1-q)(1-e^{-a}).
 \label{eq:39}
 \end{equation}

	Next, in our attempt to maximize $y$ as a function of $x$,
we may continue to hold
 \begin{equation}
 \gamma \equiv 2\pi R Z
 \label{eq:40}
 \end{equation}
fixed and hence maximize $y$ for fixed $z \equiv x/\gamma = a q$.
Then one can easily calculate that
$y = q(1-q)(1-e^{-z/q})$ has its maximum at fixed $z$ when
 \begin{equation}
 q = {e^a-1-a\over 2e^a-2-a} \approx {1\over 2}a(1-{2\over 3}a),
 \label{eq:41}
 \end{equation}
giving
 \begin{equation}
 x = \gamma a q = {\gamma a(e^a-1-a)\over 2e^a-2-a}
  \approx {1\over 2}\gamma a^2(1-{2\over 3}a)
 \label{eq:42}
 \end{equation}
and
 \begin{equation}
 y = [1-(1+a)e^{-a}]\left({e^a-1 \over 2e^a-2-a}\right)^2
   \approx {1\over 2}a^2(1-{5\over 3}a) \approx {x\over \gamma},
 \label{eq:43}
 \end{equation}
where all the approximate equalities apply for $a \ll 1$.
For fixed $\gamma$, $a$ is given implicitly as a function of $x$
by Eq. (\ref{eq:42}), and then inserting Eq. (\ref{eq:43})
for $y$ into Eq. (\ref{eq:21}) for $S_{\rm vN}(y)$
gives the entropy (so far maximized over $\beta$ and $q$)
of the density matrix (\ref{eq:17})
explicitly as a function of $a = <\!0|h^2|0\!>$
and hence implicitly as a function of $x$.
In fact, for $x \ll \gamma$, we get the asymptotic relation
 \begin{equation}
 S_{\rm vN}(y) \sim {x\over \gamma}\left(\ln{\gamma\over x} + 1 \right),
 \label{eq:44}
 \end{equation}
which of course exceeds $x$ for sufficiently small $x$.

	As the final step in the maximization of the
von Neumann entropy $S_{\rm vN}(y)$ of a density matrix
of the particular form (\ref{eq:17})
for fixed $x \equiv 2\pi R E = \gamma z = \gamma a q$,
we note that maximizing $y$ (and hence $S_{\rm vN}(y)$)
for fixed $x$ is equivalent to minimizing $x$ for fixed $y$.
Therefore, we need to minimize
$\gamma \equiv 2\pi R Z \equiv 2\pi R Y/X$.
By looking at Eqs. (\ref{eq:27}) and (\ref{eq:28}) for $X$
and $Y$, we see that the ratio $Z \equiv Y/X$
is invariant under any constant rescaling of the functions
$F(\mathbf{x})$ and $G(\mathbf{x})$
that appear as smearing functions
for $\phi(t\!=\!0,\mathbf{x})$ and $\dot{\phi}(t\!=\!0,\mathbf{x})$
in the defining Eq. (\ref{eq:23})
for the linear hermitian field operators
$\chi$ and $h = \alpha \chi$ [now that we have set $\beta = 0$
to drop the nonlinear term for $h$ in Eq. (\ref{eq:24})].
The quantity $\gamma$ is also
invariant under a rescaling of the radius $R$
if $F(\mathbf{x})$ and $F(\mathbf{x})$ depend
only on $\mathbf{x}/R$ and on some overall constant factor
that can depend on $R$.

	Minimizing $\gamma = 2\pi R Y/X$
is thus equivalent to maximizing $X$ for fixed $Y$,
which by Eq. (\ref{eq:28}) is half the integral of the sum of the
squares of $F(\mathbf{x})$ and of the gradient
of $G(\mathbf{x})$.  Because the double integral (\ref{eq:27})
for $X$ is also quadratic in $F(\mathbf{x})$ and in
$\mathbf{\nabla}G(\mathbf{x})$
but has a positive-definite nonlocal kernel,
maximizing $X$ at fixed $Y$ is best done
with fairly smooth functions $F(\mathbf{x})$
and $G(\mathbf{x})$.
In particular, if $F(\mathbf{x})$ is expanded
in spherical harmonics, one can readily see
that the maximum is obtained by keeping
only the spherically symmetric ($\ell = 0$) terms.
Also, since it is the dot product of
$\mathbf{\nabla}G(\mathbf{x})$
and $\mathbf{\nabla}G(\mathbf{y})$
that enters into Eq. (\ref{eq:27}),
which generically dilutes its contribution
relative to that of $F(\mathbf{x})F(\mathbf{y})$
for the same values of the integrals of
$|F(\mathbf{x})|^2$ and of $|\nabla G(\mathbf{x})|^2$
in Eq. (\ref{eq:28}),
one can readily see that the maximum for $X$
at fixed $Y$ is obtained by setting
$G(\mathbf{x}) = 0$, as well as choosing a
spherically symmetric
$F(\mathbf{x}) = f(|\mathbf{x}|/R)/R^2$
as given by Eq. (\ref{eq:31}).
Then one gets that
 \begin{equation}
 \gamma = {4\pi^2 \int_0^1 u^2 du f^2(u) \over
         \int_0^1 du \int_0^1 dv \:
 	 2uv\ln{\left|{u+v\over u-v}\right|}f(u)f(v)}.
 \label{eq:45}
 \end{equation}
 
	One then sees that
the minimum value for $\gamma$ is
 \begin{equation}
 \gamma = {4\pi^2 \over \lambda},
 \label{eq:46}
 \end{equation}
where $\lambda$ is the largest eigenvalue
of the weakly singular linear Fredholm
integral equation of the third kind,
 \begin{equation}
 \int_0^1 dv \: 2\ln{\left|{u+v\over u-v}\right|}w(v) = \lambda w(u)
 \label{eq:46b}
 \end{equation}
for $0 \leq u \leq 1$,
with $w(u) = u f(u)$ being the eigenfunction.

	For $F(\mathbf{x}) = f(|\mathbf{x}|/R)/R^2$
to be a smooth function of $\mathbf{x}$,
$f$ should be a smooth even function of
its argument $u \equiv |\mathbf{x}|/R$.
This means that $w(u)$
should be a smooth odd function of $u$,
so we can expand it as an infinite sum
of odd Legendre polynomials $P_{2m-1}(u)$,
 \begin{equation}
 w(u) = \sum_{m=1}^{\infty}c_m P_{2m-1}(u).
 \label{eq:47}
 \end{equation}
This expansion converts the integral eigenvalue
Eq. (\ref{eq:46b}) into the matrix eigenvalue equation
 \begin{equation}
 \sum_{n=1}^{\infty}A_{mn}c_n = \lambda \sum_{n=1}^{\infty}B_{mn}c_n,
 \label{eq:48}
 \end{equation}
where the matrix components are
 \begin{eqnarray}
 A_{mn} &=& \int_0^1 du \int_0^1 dv \:
 	 2\ln{\left|{u+v\over u-v}\right|}P_{2m-1}(u)P_{2n-1}(v)
	 \nonumber \\
	&=& {2\over [1-4(m-n)^2](m+n)(m+n-1)}
 \label{eq:49}
 \end{eqnarray}
and
 \begin{equation}
 B_{mn} = \int_0^1 du \int_0^1 dv P_{2m-1}(u)P_{2n-1}(v)
	= {\delta_{mn}\over 4m-1}.
 \label{eq:50}
 \end{equation}
 
	[Actually, I cheated slightly in obtaining
the explicit expression above for the matrix components $A_{mn}$.
I calculated $A_{11}=1$ by hand, but when I tried to calculate
the general $A_{mn}$, I got finite sums that I did not
readily see how to simplify.
Therefore, I resorted to Maple.
I did not quickly see how to get it
to give me a simple general expression
for $A_{mn}$ either, but in one afternoon I was able
to get it to give me all the values for $m < 10$, $n < 10$
(45 different terms, since $A_{mn}=A_{nm}$).
The form of these terms was sufficiently simple
that part way through their rather slow evaluation I was able
to deduce the simple expression given in Eq. (\ref{eq:49}),
which indeed fit all 45 terms.
So although I have not bothered to find a rigorous proof
that Eq. (\ref{eq:49}) is correct for all $m$ and $n$
not both smaller than 10,
the fact that it is a very simple formula that works for
all 45 smaller values strongly suggests that
it is exact for all values of $m$ and $n$.
I could say that the proof is left as an exercise
for the reader.]

	Maple readily solved the matrix eigenvalue
Eq. (\ref{eq:48}) for various truncations of the infinite
matrices $A_{mn}$ and $B_{mn}$.  For example, 40-digit
precision for $70\times 70$, $80\times 80$, $90\times 90$,
$100\times 100$, $110\times 110$, and $200\times 200$
truncations all gave the largest eigenvalue
agreeing to 13 digits:
 \begin{equation}
 \lambda \approx 3.132010216749.
 \label{eq:51}
 \end{equation}
A 20-digit calculation of the $60\times 60$ case gave the
last digit 8 instead of 9 but was used to get the
following approximate expansion of the eigenfunction
corresponding to the largest eigenvalue:
 \begin{eqnarray}
 w(u) \approx &+&P_1(u) -0.3968319408 P_3(u) + 0.0102661635 P_5(u)
  \nonumber \\
 &-&0.0070631137 P_7(u) -0.0032552106 P_9(u) -0.0018849293 P_{11}(u)
  \nonumber \\
 &-&0.0011814233 P_{13}(u) -0.0007878354 P_{15}(u) -0.0005510316 P_{17}(u) \nonumber \\
 &-&0.0004002345 P_{19}(u) -0.0002997366 P_{21}(u) -0.0002302203 P_{23}(u) \nonumber \\
 &-&0.0001806208 P_{25}(u) -0.0001442924 P_{27}(u) -0.0001170806 P_{19}(u) \nonumber \\
 &+& {\rm terms \ with \ coefficients \ less \ than \ 0.0001}.
 \label{eq:52}
 \end{eqnarray}

	One can notice that only
$P_1(u) = u$ and $P_3(u) = -1.5 u + 2.5 u^3$
give large contributions to the eigenfunction,
so one can get a fairly accurate estimate of
the largest eigenfunction by taking even just the
$2\times 2$ truncation of the matrices,
which gives the eigenvalue
 \begin{equation}
 \lambda_2 = {5(15+\sqrt{57})\over 36} \approx 3.131921449343,
 \label{eq:53}
 \end{equation}
which is smaller than the actual largest eigenvalue for the
infinite matrices by less than one part in 35\,283.
An even simpler, but rather {\it ad hoc}, approximation
is to change the coefficient of the $P_3(x)$ term above
to 0.4 and drop all the higher terms.
Dividing this trial function for $w(u)$ by $u$
(and multiplying by 8 to avoid fractions in the answer)
gives $f(u) = 8 - 5 u^2$,
which may be inserted into Eqs. (\ref{eq:45}) and (\ref{eq:46})
to give another estimate for $\lambda$,
 \begin{equation}
 \lambda_{\rm est} = {1757\over 561} \approx 3.131907308378,
 \label{eq:54}
 \end{equation}
which has almost 16\% more error than $\lambda_2$,
though this is still only a tiny error, being
smaller than the actual largest eigenvalue for the
infinite matrices by less than one part in 30\,434.
Even the very crude constant trial function
for $f(u)$ gives an eigenvalue estimate, $\lambda_{\rm crude} = 3$,
that is smaller than the actual largest eigenvalue
for the infinite matrices by only about 4.215\%,
or less than one part in 23.

	Using the approximation Maple gave for $\lambda$,
the largest eigenvalue of the infinite matrices,
Eq. (\ref{eq:46}) then gives
 \begin{equation}
 \gamma = {4\pi^2 \over \lambda} \approx 12.604817632215,
 \label{eq:55}
 \end{equation}
which can be used in Eq. (\ref{eq:44}) to get
the asymptotic behavior of $S_{\rm vN}(x)$ at sufficiently small $x$.
One can then see that this gives $S_{\rm vN}(x) > x$ for
 \begin{equation}
 x < \gamma e^{1-\gamma} \approx 0.000115,
 \label{eq:56}
 \end{equation}
or alternatively for 
 \begin{equation}
 S_{\rm vN} < \gamma e^{1-\gamma} \approx 0.000115.
 \label{eq:57}
 \end{equation}
Thus if the dimensionless energy, $x \equiv 2\pi R E$,
and the von Neumann entropy, $S = S_{\rm vN} \equiv - tr\rho\ln\rho$,
are sufficiently small, then with the definitions used here
for these quantities, they can violate Bekenstein's
conjectured entropy bound (\ref{eq:1}), $S \leq x$,
though admittedly the range of $x$ and $S$ for which
this happens is very narrow.

	We may now use the value of $\gamma$, given by Eq.
(\ref{eq:55}), in Eqs. (\ref{eq:21}), (\ref{eq:42}), and (\ref{eq:43})
to get $B \equiv S_{\rm vN}/x$ as a precise implicit function purely of
$x$, or, alternatively, to get both $x$ and $B$ as explicit functions of
$a = <\!0|h^2|0\!>$. Of course, this is merely for one simple example of
a one-parameter family of mixed states given by Eq. (\ref{eq:17}), with
$q$ given by Eq. (\ref{eq:41}) and $|\psi\!>$ given by Eq. (\ref{eq:14})
with $\beta = 0$ so $h = \alpha \chi$ and with $\chi$ given by Eq.
(\ref{eq:23}) with $G(\mathbf{x}) = 0$ and Eq. (\ref{eq:31}) giving
$F(\mathbf{x}) = f(|\mathbf{x}|/R)/R^2$ with $w(u) = u f(u)$ being an
eigenvector corresponding to the largest eigenvalue, $\lambda$, of the
homogeneous linear Fredholm integral equation (\ref{eq:46b}). Therefore,
it is not likely to give the maximum possible $B$ as a function of $x$,
which was called $B_{\rm N}(x)$ in Eq. (\ref{eq:7c}). However, this
$B(x)$ does give at least a lower bound on $B_{\rm N}(x)$ for a single
massless scalar field.

\section{Conjectures for Entropy Bounds of Vacuum-Outside-R States}

	I would conjecture that asymptotically at small
$x \equiv 2 \pi R E$,
the density matrix (\ref{eq:17}), with all the entropy-maximization
procedures given above for a density matrix of this form,
gives $B(x)$ that does asymptotically approach the unknown global
maximum function $B_{\rm N}(x)$ for a single massless scalar field.
Therefore, if we divide $x$ into the asymptotic form of $S_{\rm vN}$
for small $x$ that is given by Eq. (\ref{eq:44}),
I would conjecture that this gives the asymptotic
form of the true upper bound, $B_{\rm N}(x)$, for very small $x$,
 \begin{equation}
 B_{\rm N}(x) \sim {1\over \gamma}\left(\ln{\gamma\over x} + 1 \right),
 \label{eq:58}
 \end{equation}
with Eq. (\ref{eq:55}) giving $\gamma \approx 12.604817632215$. 
One might expect a similar formula for other free
massless fields, though perhaps each with a different value of $\gamma$.

	When $x$ is not small, it is certainly not the case that the
density matrix of fixed $x$ needed to maximize the entropy is
approximately of the simple rank-two form given by Eq. (\ref{eq:17}). 
One would surely need a more general vacuum-outside-$R$ state, with a
density matrix obeying Eq. (\ref{eq:2}) (giving vacuum expectation
values for all operators not in causal contact with the ball $r \leq R$
at $t = 0$), such as that given by Eq. (\ref{eq:13}), most likely with
an infinite sum of terms and an infinite rank. I do not know how to
proceed toward finding such a density matrix obeying Eq. (\ref{eq:2})
that would maximize $B(x) \equiv S/x$ at finite $x$ that is neither
asymptotically small or large. However, one might try using in Eq.
(\ref{eq:13}) $h_i$'s that have the form given in Eq. (\ref{eq:24}),
with the $F(\mathbf{x})$'s and $G(\mathbf{x})$'s of Eq. (\ref{eq:23})
being suitable eigenfunctions of the three-dimensional version of the
integral equation (\ref{eq:46b}). Even this wide class of examples may
not be sufficient, since one could imagine instead constructing the
Hermitian operators $h_i$ from smeared functions of the field
$\phi(t\!=\!0,\mathbf{x})$ and of its time-derivative (or conjugate
momentum) $\dot{\phi}(t\!=\!0,\mathbf{x})$ that are not merely linear as
is the $\chi$ given by Eq. (\ref{eq:23}). Going to nonlinear Hermitian
operators (other than the relatively simple $\chi^2$ considered above)
leads to such a wealth of possibilities that I do not presently know how
to proceed to obtain a true maximum for $B(x)$ at fixed finite $x$, the
postulated function $B_{\rm N}(x)$.

	In the more usual case of boundary conditions on the field,
quantum states obeying these boundary conditions may be coherently
superposed (i.e., the corresponding wavefunctions added, not just the
density matrices) to get other states that also obey the boundary
conditions. Then one can look for superpositions that diagonalize the
Hamiltonian (i.e., energy eigenstates). From these, one can form a Gibbs
ensemble to maximize the von Neumann entropy at a fixed expectation
value of the energy.

	However, for the vacuum-outside-$R$ states considered here,
coherent superpositions of pure vacuum-outside-$R$ states are
generically not vacuum-outside-$R$ states. (Of course, positive-weight
combinations of vacuum-outside-$R$ density matrices are still
vacuum-outside-$R$ density matrices when normalized, since the
vacuum-outside-$R$ condition, that all expectation values outside the
ball $r\leq R$ at $t=0$ are the same as the vacuum, is homogeneous and
linear in the density matrix, though not in the wavefunction.)
Therefore, the procedure for diagonalizing the Hamiltonian for such
states fails.

	Indeed, one can see that none of the vacuum-outside-$R$ states,
except for the vacuum itself, can be an energy eigenstate.  This is
because any state which is non-vacuum in a finite region at some time
will inevitably have that region spread with time. For fields with
linear field equations, the perturbations of the field itself will
spread. But even for fields with self-coupling which allow classical
field solitons that do not spread with time, any quantum state of the
field which is non-vacuum at some time will inevitably have that region
spread with time as a result of the quantum uncertainty principle. For
example, suppose that there is some definition of the location and
momentum of the soliton, such that the velocity of the location is
proportional to the momentum.  Then the position-momentum uncertainty
principle will prevent one from having that the location remain, with
certainty, within any finite region for an infinite amount of time; the
quantum uncertainty of the position, if initially confined to a finite
region, will inevitably spread to extend all over space. Therefore, the
confined configuration cannot be stationary and hence cannot be an
energy eigenstate.

	One can test my conjecture that Eq. (\ref{eq:58})
is the correct asymptotic form of the true upper bound
on the entropy per $x \equiv 2\pi RE$ by examining
some other simple density matrices of the form given by
Eq. (\ref{eq:13}) that allow explicit evaluation of the
von Neumann entropy.  For example, the rank-three density matrix
 \begin{equation}
 \rho = (1-q)|0\!><\!0|+(q/2)e^{i\alpha\chi}|0\!><\!0|e^{-i\alpha\chi}
 	+(q/2)e^{-i\alpha\chi}|0\!><\!0|e^{i\alpha\chi}
 \label{eq:59}
 \end{equation}
has $z\equiv x/\gamma = aq$, just like rank-two density matrix
(\ref{eq:17}) when $\beta = 0$,
and it has the three nonzero eigenvalues
 \begin{eqnarray}
 p_1 &=& {1\over 2}[1-{1\over 2}q(1-e^{-2a})]
 +{1\over 2}\sqrt{[1-{1\over 2}q(1-e^{-2a})]^2-2q(1-q)(1-e^{-a})^2}
 \nonumber \\
 &\approx& 1-q(1-e^{-a})+{1\over 2}q^2(1-e^{-a})^2
 \approx 1 - qa + {1\over 2}(q+q^2)a^2,
 \label{eq:60}
 \end{eqnarray}
 \begin{eqnarray}
 p_2 &=& {1\over 2}q(1-e^{-2a})
 \nonumber \\
 &\approx& qa(1-a),
 \label{eq:61}
 \end{eqnarray}
and
 \begin{eqnarray}
 p_3 &=& {1\over 2}[1-{1\over 2}q(1-e^{-2a})]
 -{1\over 2}\sqrt{[1-{1\over 2}q(1-e^{-2a})]^2-2q(1-q)(1-e^{-a})^2}
 \nonumber \\
 &\approx& {1\over 2}q(1-q)(1-e^{-a})^2
 \approx {1\over 2}q(1-q)a^2,
 \label{eq:62}
 \end{eqnarray}
where the approximate equations apply for very small $a\equiv \alpha^2
X$. If one chooses $q$ to maximize the von Neumann entropy of this mixed
state,
 \begin{equation}
 S_{\rm vN} \equiv - tr(\rho\ln\rho)
  = -p_1\ln{p_1}-p_2\ln{p_2}-p_3\ln{p_3},
 \label{eq:63}
 \end{equation}
for very small $a$ (equivalently, very small $z$),
one gets that $q \approx (1/3)(1+4a/3)$,
so $z = aq \approx (a/3)+4(a/3)^2$, which may be inverted to give
$a \approx 3z(1-4z)$ and $q \approx (1/3)(1+4z)$.
This then gives
 \begin{equation}
 S_{\rm vN} \approx z[1+z-(1-z)\ln{z}]
 \sim {x\over \gamma}\left(\ln{\gamma\over x} + 1 \right),
 \label{eq:64}
 \end{equation}
which has the same asymptotic form for small $x$ as Eq. (\ref{eq:44})
for the rank-two density matrix (\ref{eq:17}).

	Another example would be to consider the rank-five
density matrix
 \begin{eqnarray}
 \rho = (1-q)|0\!><\!0|
 &+& (q/4)e^{i\alpha\chi+i\beta\chi^2}|0\!><\!0|e^{-i\alpha\chi-i\beta\chi^2}
	\nonumber \\
 &+&(q/4)e^{i\alpha\chi-i\beta\chi^2}|0\!><\!0|e^{-i\alpha\chi+i\beta\chi^2}
	\nonumber \\
 &+& (q/4)e^{-i\alpha\chi+i\beta\chi^2}|0\!><\!0|e^{i\alpha\chi-i\beta\chi^2}
	\nonumber \\
 &+&(q/4)e^{-i\alpha\chi-i\beta\chi^2}|0\!><\!0|e^{i\alpha\chi+i\beta\chi^2},
 \label{eq:65}
 \end{eqnarray}
which gives $z\equiv x/\gamma = q(a+b)$, where, as above,
$a \equiv \alpha^2 X$ and $b \equiv 4 \beta^2 X^2$.
Even though the eigenvalue equation is now a fifth-order polynomial
equation, it appears that one may be able to use the symmetries
of the problem to find the eigenvectors and eigenvalues
explicitly, as functions of $q$, $a$, and $b$,
without requiring any roots higher than square roots.
However, this seems to be messier than is worth doing here,
so it shall be left as another exercise for the reader.

	Nevertheless, one can show that when $a \ll 1$ and $b \ll 1$,
there is one eigenvalue near unity, one near $qa$, one near $qb/2$,
and the remaining two are smaller by factors of the order of $z = q(a+b)$.
Therefore, in this limit only the three largest eigenvalues
contribute significantly to the von Neumann entropy, giving
 \begin{equation}
 S_{\rm vN} \approx
  -(1-qa-qb/2)\ln{(1-qa-qb/2)}-qa\ln{(qa)}-(qb/2)\ln{(qb/2)}.
 \label{eq:66}
 \end{equation}
When this is maximized at fixed $z = q(a+b)$, one finds that
the first three eigenvalues need to be approximately
in a geometric series (as, e.g., are all the eigenvalues
of the thermal density matrix for an harmonic oscillator), giving
$qb/2 \approx (qa)^2(1+qa)$.
Solving for $qa$ and $qb$ in terms of $z$ and inserting
this back into Eq. (\ref{eq:66}) gives the maximum von Neumann entropy
 \begin{equation}
 S_{\rm vN} \approx z[1+{1\over 2}z-\ln{z}]
 \sim {x\over \gamma}\left(\ln{\gamma\over x} + 1 \right)
 \label{eq:67}
 \end{equation}
for the rank-five density matrix (\ref{eq:65}) at fixed tiny $z \equiv
x/\gamma$. This entropy is just slightly larger, by an amount roughly
$z^2[\ln{(1/z)}-(1/2)]$, than the corresponding maximum entropy
(\ref{eq:64}) for the rank-three density matrix (\ref{eq:59}) at tiny $z
= x/\gamma$, but it has the same asymptotic limit (given after the
$\sim$ sign).

	Therefore, although of course these three simple examples of
finite-rank density matrices do not begin to exhaust the infinite set of
possibilities, they give some support to the conjecture given above that
Eq. (\ref{eq:58}) is the correct asymptotic form of the upper bound for
$S/(2\pi R E)$ when the denominator of this expression, $x \equiv 2\pi R
E$, is much smaller than unity. 

	In the opposite limit, when $x \equiv 2\pi R E$ is much greater
than unity, we would expect that, at least for a scale-invariant field
so that the energy $E$ is large with respect to all relevant parameters
with the same dimension ($1/R$ being the only relevant one if the field
does not have a rest mass or other parameter setting a higher energy
scale), the maximum entropy is given by Eq. (\ref{eq:9}) for
high-temperature thermal radiation, giving
 \begin{equation}
 B_{\rm N}(x) \approx {\beta\over x^{1/4}},
 \label{eq:89}
 \end{equation}
with now
 \begin{equation}
 \beta = \left[{2^6\over 3^6 5}(n_b + {7\over 8}n_f)\right]^{1\over 4},
 \label{eq:90}
 \end{equation}
no longer the $\beta$ of Eq. (\ref{eq:24}) that we have subsequently set
to zero to maximize $B(x)$ for the density matrix (\ref{eq:17}). It is
tempting to combine this asymptotic formula for $x \gg 1$ with the
asymptotic formula (\ref{eq:58}) conjectured above for $x \ll 1$ to
conjecture that a reasonably good approximate formula for $B_{\rm
N}(x)$, as a function of any $x \equiv 2\pi R E$, for a quantum field
theory with a given set of massless fields, is
 \begin{equation}
 B_{\rm N}(x)
  \simeq {4\over\gamma}\ln{\left(1+{\beta\gamma\over 4 x^{1/4}}\right)},
 \label{eq:91}
 \end{equation}
where the constants $\beta$ and $\gamma$ would depend upon
the massless fields in the theory.
For the single massless real scalar field that has been considered here,
$n_b = 1$ and $n_f = 0$, so Eq. (\ref{eq:24}) would give
 \begin{equation}
 \beta = \left({2^6\over 3^6 5}\right)^{1\over 4}
  \approx 0.364016115028,
 \label{eq:92}
 \end{equation}
and Eq. (\ref{eq:55}) has already given $\gamma \approx 12.604817632215$
for the single real massless scalar field.

	Another way to state this conjecture is to write
 \begin{equation}
 B_{\rm N}(x)
  = {4\over\gamma}\ln{\left(1+{\beta\gamma\over 4 x^{1/4}}\right)} C(x),
 \label{eq:93}
 \end{equation}
where $C(x)$ is a correction factor yet to be found,
and then conjecture that $C(x)$ tends asymptotically to unity
for both very small and very large $x$,
and perhaps further to conjecture that $C(x)$ is always relatively
close to unity (e.g., say within a factor of two).
This conjecture would then imply a conjectured entropy bound,
 \begin{equation}
 S_{\rm vN} \leq {8\over\gamma}\ln{\left(1+{\beta\gamma
 \over 4 (2\pi R E)^{1/4}}\right)} 2\pi R E.
 \label{eq:94}
 \end{equation}

	An improved formula might be to write
 \begin{equation}
 B_{\rm N}(x) = {4\over\gamma}\ln{\left({1+Ax^{-1/4}+Bx^{-1/2}
 \over 1+Cx^{-1/4}}\right)} \tilde{C}(x)
 \label{eq:95}
 \end{equation}
with $B/C=(e\gamma)^{1/4}$ to fit the final 1 in
the asymptotic formula (\ref{eq:58}) for $x \ll 1$,
$A-C=\beta\gamma/4$ to fit
the asymptotic formula (\ref{eq:89}) for $x \gg 1$,
and $2B-A^2+C^2=\gamma\delta/2$ to fit the following
two-term improvement to Eq. (\ref{eq:89}):
 \begin{equation}
 B_{\rm N}(x) \approx {\beta\over x^{1/4}} + {\delta\over x^{1/2}}.
 \label{eq:96}
 \end{equation}
I have not tried to work out what $\delta$ is.
It would be straightforward to calculate, if it were the same
as for the thermal state of a massless scalar field
inside a sphere with Dirichlet boundary conditions
on the field at the boundary $r=R$,
but it is not obvious to me whether or not it is the same.

	Since Eq. (\ref{eq:95}) with the correction factor
$\tilde{C}(x)$ omitted should give a better asymptotic fit
to $B_{\rm N}(x)$ than Eq. (\ref{eq:91}),
I would expect that $\tilde{C}(x)$ would generally be closer to unity
than the corresponding correction factor $C(x)$ of Eq. (\ref{eq:93})
[not to be confused with the coefficient $C$ in Eq. (\ref{eq:95})].
But whether this is true over the entire infinite range of $x$
remains to be seen.

	For free massive quantum fields, for fixed entropy one would
expect that the energy would have to be higher, so an upper bound for a
set of free massless quantum fields should also give an upper bound for
a corresponding set of free massive quantum fields. Therefore, I would
conjecture that for any given free quantum field theory, one can find a
$\beta$ and $\gamma$ (presumably with $\beta$ obeying Eq. (\ref{eq:92}),
and $\gamma$ some combination of eigenvalues of the appropriate integral
equations) such that the inequality (\ref{eq:94}) holds with that value
of $\beta$ and $\gamma$. The conjecture might even be true for any
reasonable interacting quantum field theory that is causal in the sense
defined above, though then one might need a different value of $\beta$.

\section{Possibilities for Trying to Retain Bekenstein's Proposed Bound}

	Returning to a consideration of how Bekenstein's
proposed bound (\ref{eq:1})
fits with the results derived and conjectured here,
one first notes that the results here violate (\ref{eq:1})
for the von Neumann entropy $S_{\rm vN}$ of a vacuum-outside-$R$
mixed state with sufficiently small energy expectation value $E$,
e.g., for $x \equiv 2\pi RE < 0.000115$ in the example above.
However, even if one accepts the use of vacuum-outside-$R$ states
for defining a finite size $R$, one might still object
that the Bekenstein bound is not intended to be applied
to the definition of $E$ and/or $S$ being used here.

	For example, Schiffer and Bekenstein
\cite{SB1}
refer to ``quantum states accessible to the field system
with energy up to and including $E$.''
It could be objected that since the vacuum-outside-$R$ states
considered above are not energy eigenstates,
they are actually composed of states with energy both lower
and higher than the energy expectation value that I have
used as the definition of $E$.
If one takes $E$ to be the energy of one of the energy eigenstates
that is sufficiently higher than the expectation value,
then the Bekenstein bound (\ref{eq:1}) may be obeyed
even in the examples I have given above that violate the bound
when $E$ is taken to be the energy expectation value.

	But if one takes this approach, it is hard
to see how to give any content to the proposed bound
for the vacuum-outside-$R$ states.
Presumably not only is it the case that any
vacuum-outside-$R$ state is not an energy eigenstate
(since it is not stationary), but also it is surely the
case that if any vacuum-outside-$R$ state
is decomposed into energy eigenstates,
it will include energy eigenvalues of arbitrarily
large value.  However, using an arbitrarily large
value of $E$ in the bound (\ref{eq:1}) makes it trivial,
entropy less than or equal to infinity.
Therefore, for the bound to have any content,
we need to have a definition of $E$ that gives finite values.
The definition given by Eq. (\ref{eq:3}) above,
$E \equiv tr(H\rho)$, is surely the simplest,
though others could be proposed.

	For example, one could propose instead that
for a vacuum-outside-$R$ density matrix of the form
(\ref{eq:13}), $E$ could be defined as the maximum
value of the expectation value of the Hamiltonian $H$
in any of the normalized states $e^{i h_i}|0\!><\!0|e^{-i h_i}$
whose sum, weighted by the $q_i$'s, forms $\rho$.
However, using this definition would not avoid
violations of Bekenstein's bound (\ref{eq:1}).
For example, one could use the density matrix (\ref{eq:59})
with $q=1$ so that the first term vanishes,
and then the remaining two terms are of the form (\ref{eq:13})
with $h_1 = \alpha\chi$ and $h_2 = -\alpha\chi$,
and with $q_1=q_2=1/2$.  Each of the two nonzero terms
of the density matrix then gives the same energy
expectation value, and one can calculate that
for small $z=x/\gamma=a\equiv \alpha^2 X$ one gets
 \begin{equation}
 S_{\rm vN} \approx z[1-{1\over 2}z-(1-z)\ln{z}]
 \sim {x\over \gamma}\left(\ln{\gamma\over x} + 1 \right),
 \label{eq:97}
 \end{equation}
again violating Bekenstein's bound (\ref{eq:1})
for $x < 0.000115$.

	Yet another proposal would be that $E$
be defined as $E_{\rm max}$,
the maximum expectation value
of the Hamiltonian in all of the eigenstates
of the density matrix.
In the example just discussed,
 \begin{equation}
 \rho = p_1|1\!><\!1|+p_2|2\!><\!2|
 \label{eq:98}
 \end{equation}
with $p_1 = (1+e^{-2a})/2$, $p_2 = (1-e^{-2a})/2$,
and orthonormal eigenstates
 \begin{equation}
 |1\!>={\cos{(\alpha\chi)}|0\!>\over \sqrt{p_1}}
 \label{eq:99}
 \end{equation}
and
 \begin{equation}
 |2\!>={\sin{(\alpha\chi)}|0\!>\over \sqrt{p_2}}.
 \label{eq:100}
 \end{equation}
Then by using Eq. (\ref{eq:30}),
one can calculate not only that
 \begin{eqnarray}
 <\!0|e^{-ih_1}He^{ih_1}|0\!>
 \equiv <\!0|e^{-i\alpha\chi}He^{i\alpha\chi}|0\!>
 \!\!\!&=&\!\!\! <\!0|e^{-ih_2}He^{ih_2}|0\!>
 \equiv <\!0|e^{i\alpha\chi}He^{-i\alpha\chi}|0\!>
 \nonumber \\
 \!\!\!&=&\!\!\! \alpha^2 Y \equiv \alpha^2 X Z \equiv a Z
 \label{eq:101}
 \end{eqnarray}
as given by Eq. (\ref{eq:26}), but also
 \begin{eqnarray}
 <\!0|e^{-ih_1}He^{-ih_2}|0\!>
 \equiv <\!0|e^{-i\alpha\chi}He^{-i\alpha\chi}|0\!>
 \!\!\!&=&\!\!\! <\!0|e^{ih_2}He^{ih_1}|0\!>
 \equiv <\!0|e^{i\alpha\chi}He^{i\alpha\chi}|0\!>
 \nonumber \\
 \!\!\!&=&\!\!\! a Z e^{-a}.
 \label{eq:102}
 \end{eqnarray}
From these results and from the form of the orthonormal
eigenstates given by Eqs. (\ref{eq:99}) and (\ref{eq:100}),
one readily obtains
 \begin{equation}
 H_{11} \equiv <\!1|H|1\!>={aZ(1-e^{-a})\over 1+e^{-2a}}
 \approx {1\over 2}a^2 Z
 \label{eq:103}
 \end{equation}
and
 \begin{equation}
 H_{22} \equiv <\!2|H|2\!>={aZ\over 1-e^{-a}}
 \approx Z
 \label{eq:104}
 \end{equation}
Then if one takes $E_{\rm max}$,
the larger of $H_{11}$ and $H_{22}$,
namely $H_{22}$, as the definition of $E$,
one sees that it has the positive lower limit $Z=\gamma/(2\pi R)$
as one takes $a$ (and hence $S_{\rm vN}$) to zero,
so with this definition one does not get a violation
of Bekenstein's bound (\ref{eq:1}) in this example.
In particular, for this example
 \begin{eqnarray}
 B_{E_{\rm max}} \equiv {S_{\rm vN}\over 2\pi R E_{\rm max}}
 &=& {1-e^{-a}\over \gamma a}
   \{-{1\over 2}(1+e^{-2a})\ln{[{1\over 2}(1+e^{-2a})]}
   \nonumber \\
 &-& {1\over 2}(1-e^{-2a})\ln{[{1\over 2}(1-e^{-2a})]}\} < 1.
 \label{eq:105}
 \end{eqnarray}
Whether this definition of $B_{E_{\rm max}}$
always gives a result less than unity
for all vacuum-outside-$R$ states,
thus agreeing with Bekenstein's bound,
remains to be proven, but the extremely meagre
evidence that I have does not seem to contradict
this conjecture.

	Another objection that might be made
against the violations of Bekenstein's bound (\ref{eq:1})
using vacuum-outside-$R$ states to define $R$,
using the expectation value of the Hamiltonian
given by Eq. (\ref{eq:3}) as the definition of the energy $E$,
and using the von Neumann entropy given by Eq. (\ref{eq:4})
as the definition of the entropy $S$,
is to demand that the entropy instead be given by
a microcanonical ensemble rather than
by Eq. (\ref{eq:4}) applied to any mixed state.
In other words, instead of allowing
a generic vacuum-outside-$R$ mixed state
or density matrix $\rho$ obeying Eq. (\ref{eq:2})
for all operators $O$ completely confined to
the region $D$, $r > R + |t|$,
one might propose that Bekenstein's conjectured
bound should only be applied to density matrices
made up of equal mixtures of $n$ orthogonal
vacuum-outside-$R$ pure states.
(These are rank-$n$ density matrices
with precisely $n$ nonzero eigenvalues,
all equal to $1/n$, and with the corresponding
set of $n$ orthonormal eigenvectors all being
vacuum-outside-$R$ pure states.)
The entropy of such a density matrix
whose nontrivial part is proportional
to the identity matrix in the $n$
nontrivial dimensions is then $S=\ln{n}$.

	Again, I do not have evidence that
Bekenstein's conjectured bound (\ref{eq:1})
is violated for such restricted
vacuum-outside-$R$ density matrices.
However, it is a rather severe limitation
to restrict the discussion to such a small
subset of vacuum-outside-$R$ density matrices,
a subset of measure zero in the space
of all such density matrices.
Furthermore, it appears rather difficult
to find many explicit examples of
precisely orthogonal vacuum-outside-$R$
pure states.

	For example, for each fixed choice of the
two functions $F(\mathbf{x})$ and $G(\mathbf{x})$
in Eq. (\ref{eq:23}), the constants $\alpha$
and $\beta$ in Eq. (\ref{eq:24}) give a two-parameter
family of vacuum-outside-$R$ pure states
of the form $e^{i\alpha\chi+i\beta\chi^2}|0\!>$,
and then Eq. (\ref{eq:25}) and its trivial
generalization gives the inner product
between any two states among this
two-parameter family.  However, none of these
inner products are zero for finite $\alpha$'s
and $\beta$'s (and hence for finite expectation
values of the energy), so none of these states
are orthogonal for fixed $F(\mathbf{x})$ and
$G(\mathbf{x})$.

	Of course, if one combines various ones
of these nonorthogonal pure state density matrices
to get a mixed density matrix and then finds
the eigenvectors of that density matrix,
they will form an orthonormal set of density
matrices, such as the set $|1\!><\!1|$
and $|2\!><\!2|$ of Eqs. (\ref{eq:99}) and (\ref{eq:100}).
However, these density matrices are not by themselves
vacuum-outside-$R$ states, but only when they are combined
with the particular eigenvalues $p_1$ and $p_2$
given just before Eq. (\ref{eq:99}).
Hence they cannot be used in a different linear combination
(e.g., with $p_1 = p_2 = 1/2$) to get a vacuum-outside-$R$
state that is an equal-weight combination of $n$
orthonormal pure-state vacuum-outside-$R$ states.

	The only explicit vacuum-outside-$R$
state orthogonal to the vacuum itself
(the trivial vacuum-outside-$R$ state)
that I have found so far is the pure state
 \begin{equation}
 |\psi\!> = \exp{\left(i\alpha e^{-\beta\chi^2}\right)}|0\!>
 \label{eq:106}
 \end{equation}
with real $\alpha$ and $\beta$
(not the same $\alpha$ and $\beta$ used elsewhere
in this paper) chosen so that
 \begin{equation}
 <\!0|\psi\!> = \sum^{\infty}_{n=0}
 {(i\alpha)^n\over n\!}(1+2n\beta X)^{-1/2} = 0.
 \label{eq:107}
 \end{equation}
Using Maple, I found a numerical solution at
 \begin{equation}
 \alpha \approx 4.727048274,\ \beta \approx 1.536994796/X.
 \label{eq:108}
 \end{equation}
I have not worked out the expectation value of the energy,
$<\!\psi|H|\psi\!>$, of this pure state,
but I suspect that it is greater than $\ln{2}/(\pi R)$,
so that the entropy of an equal mixture
of this state and of the vacuum, $\ln{2}$,
would be less than $2\pi R$ times the expectation value
of the energy in this mixed state (half the expectation
value of the energy of the pure state $|\psi\!><\!\psi|$,
since the vacuum half of the mixed state
contributes zero to the expectation value of the energy).
If so, then this example would not be a counterexample to
Bekenstein's conjectured bound (\ref{eq:1})
restricted to microcanonical ensembles
that are equal mixtures of orthogonal pure
vacuum-outside-$R$ states.

	It would be interesting to find the
lowest-energy vacuum-outside-$R$ state
orthogonal to the vacuum itself
and see whether its energy is indeed not more
than $\ln{2}/(\pi R)$,
but I do not see how to do this at present.
More generally, one would like to find, for
each positive integer $n$, the set of $n$
mutually orthogonal vacuum-outside-$R$ states
(possibly, but not necessarily, including
the vacuum itself)
such that the sum of the $n$ energy expectation values,
say $E_s$, is minimized.  Then if one finds that
$E_s \geq (n\ln{n})/(2\pi R)$ for each $n$,
then Bekenstein's conjectured bound (\ref{eq:1})
will be obeyed for these microcanonical ensembles
of vacuum-outside-$R$ states.

	One way to look for other pure vacuum-outside-$R$
states orthogonal to the vacuum
would be to choose some non-hermitian operator,
say $\kappa$, that is confined to the ball $r \leq R$ at $t=0$,
and consider the one-complex-parameter
(two-real-parameter) set of states
 \begin{equation}
 |\psi(C;\kappa)\!> = e^{iC\kappa-i\bar{C}\kappa\dagger}|0\!>
 \label{eq:116}
 \end{equation}
for the complex parameter $C$.
For a generic such $\kappa$, $<\!0|\psi(C;\kappa)\!>$
would be a complex function of $C$ (not analytic,
since both $C$ and its complex conjugate $\bar{C}$
appear in the definition of $|\psi(C;\kappa)\!>$),
and a simple parameter-counting argument suggests
that there should be discrete complex values of $C$
at which $<\!0|\psi(C;\kappa)\!>=0$,
giving a pure vacuum-outside-$R$
states orthogonal to the vacuum,
though of course for particular $\kappa$'s,
the number of such discrete solutions for $C$ may be zero.
One might extend this method to try to find
$n$ mutually orthogonal vacuum-outside-$R$ states;
this would require $n$ different operators
$\kappa_i$ and $n(n+1)/2$ complex parameters.
The obvious problem for carrying out this procedure
explicitly is that for most sets of operators $\kappa_i$,
the inner products (functions of the complex parameters)
would be difficult to evaluate.

	Perhaps a compromise to the stringent
requirement of a microcanonical ensemble of
$n$ equally-weighted orthogonal pure vacuum-outside-$R$
density matrices is simply to use the von Neumann entropy
$S_{\rm vN} \equiv - tr\rho\ln\rho$,
which equals $\ln{n}$ for a microcanonical ensemble,
but require that it be at least as large as $\ln{2}$,
the minimum nontrivial value for a microcanonical ensemble.
Then one might conjecture that the bound (\ref{eq:1})
is correct for vacuum-outside-$R$ states such that
$S=S_{\rm vN} \geq \ln{2}$.

	Alternatively, one might replace
Bekenstein's conjectured bound (\ref{eq:1}) with
the weaker conjectured bound
 \begin{equation}
 S \leq 2 \pi E R + \ln{2},
 \label{eq:117}
 \end{equation}
still using $S = S_{\rm vN} \equiv - tr\rho\ln\rho$,
$E \equiv tr(H\rho)$, and restricting to
vacuum-outside-$R$ density matrices $\rho$
obeying $tr(O\rho) = <\!0|O|0\!>$
for all operators $O$ totally confined to
the region $D$, $r > R + |t|$,
that is not in causal contact with
the ball $B$, $r \leq R$ at $t=0$
(no causal curves connecting these two regions).
(Equivalently, one may require that
$tr(O\rho) = <\!0|O|0\!>$
for all operators $O$ that commute with
all operators defined totally on the ball $B$.)

\section{Other Ways to Define a Radius R}

	So far I have been considering
only the new proposal to define $R$
by restricting to vacuum-outside-$R$ states.
However, one might ask whether there are other ways
to define a radius $R$ for a class of states
for which one is seeking a bound on the entropy $S$
as a function of R and of the energy $E$.

	One proposal that is very close to
my proposal of vacuum-outside-$R$ states
is a proposal for what might be called
stressless-outside-$R$ states,
states such that on the $t=0$
flat hyperplane of the Minkowski spacetime
that I have always been assuming so far in this paper,
the expectation value of the regularized
stress-energy tensor operator, $T_{\mu\nu}$,
is zero everywhere outside the radius $R$
(as it is everywhere for the vacuum state),
 \begin{equation}
 \tau_{\mu\nu}(\mathbf{x}) \equiv
   tr(T_{\mu\nu}(t\!=\!0,\mathbf{x})\rho) = 0
 \label{eq:118}
 \end{equation}
for all $|\mathbf{x}|>R$.
Alternatively, one might restrict to
what might be called
energyless-outside-$R$ states,
 \begin{equation}
 \varepsilon(\mathbf{x}) \equiv \tau_{00} \equiv
   tr(T_{00}(t\!=\!0,\mathbf{x})\rho) = 0
 \label{eq:119}
 \end{equation}
for all $|\mathbf{x}|>R$,
for which the expectation value
$\varepsilon(\mathbf{x})$ of
the regularized energy density operator, $T_{00}$,
at $t=0$ vanishes outside the radius $|\mathbf{x}| = R$.

	Of course, all vacuum-outside-$R$ states
are also stressless-outside-$R$ states,
and all stressless-outside-$R$ states
are also energyless-outside-$R$ states,
but I do not know whether the converses of these statements
are true.  If they are not both true,
there would exist energyless-outside-$R$ states,
and possibly also stressless-outside-$R$ states,
that are not also vacuum-outside-$R$ states.
If there is indeed a broader class of states
than vacuum-outside-$R$ states, whether
stressless-outside-$R$ states
and/or energyless-outside-$R$ states,
then the corresponding entropy maximization
function $\sigma_{\rm vN}(R,E)$
would be expected to be larger for
the broader class of states.

	One can try to define $R$ for even broader
classes of states, not by requiring that some
expectation values vanish for $r>R$ at $t=0$,
but instead by using the spatial distribution
of some quantity to define an effective radius $R$.
One way that first comes to mind is to use some
spatially-dependent real weight function $W(\mathbf{x})$
coming from the quantum state to define $R$
as an rms value of $r \equiv |\mathbf{x}|$:
 \begin{equation}
 R_W^2 = {\int d^3x W(\mathbf{x}) r^2 \over \int d^3x W(\mathbf{x})}.
 \label{eq:120}
 \end{equation}
Of course, the weight function should be such
that both the numerator and the denominator
are finite and have the same sign
(which without loss of generality
will be assumed to be positive),
at least for the class of states to be considered.

	An obvious simple choice of the weight function
is the energy density expectation value $\varepsilon(\mathbf{x})$.
Then the denominator of Eq. (\ref{eq:120}) for $R_W^2$
is the total energy, which is positive for a nontrivial state.
However, the numerator is not positive for all nontrivial states,
as one can see from the following argument:
Motivated by the state with locally negative energy density
given by Kuo and Ford
\cite{KF},
consider the state
 \begin{equation}
 |\psi\!> = \alpha|0\!> + \beta|2\!>,
 \label{eq:121}
 \end{equation}
where $|\alpha|^2 + |\beta|^2 = 1$ and
$|2\!>$ is a two-quantum state of energy
 \begin{equation}
 E_2 \equiv <\!2|H|2\!> = \int d^3x <\!2|T_{00}|2\!>
 \label{eq:122}
 \end{equation}
and with mode functions
that are sufficiently localized that
 \begin{equation}
 R_2^2 E_2 \equiv \int d^3x <\!2|T_{00}|2\!> r^2
 \label{eq:123}
 \end{equation}
is finite.
Because $T_{00}$ is the regularized (e.g., normal-ordered)
energy density operator, $<\!0|T_{00}|0\!> = 0$
and $\int d^3x <\!2|T_{00}|0\!>=\int d^3x <\!0|T_{00}|2\!>=0$.
For a generic two-quantum state $|2\!>$,
 \begin{equation}
 C \equiv \int d^3x <\!2|T_{00}|0\!> r^2
 \label{eq:124}
 \end{equation}
will be a nonzero complex number.
Then Eq. (\ref{eq:120}) for
$W(\mathbf{x}) = \varepsilon(\mathbf{x})$ gives
 \begin{equation}
 R_W^2 = {\int d^3x <\!\psi|T_{00}|\psi\!> r^2
 \over \int d^3x <\!\psi|T_{00}|\psi\!> r^2}
	= R_2^2 + \Re\left({2C\over E_2}{\alpha\over\beta}\right).
 \label{eq:125}
 \end{equation}
Since $\alpha/\beta$ can be an arbitrary complex number,
$R_W^2$ can take any real value if $C\neq 0$,
including zero and negative values.
Even if one restricted to states for which $R_W^2$ is positive,
this quantity can be made arbitrarily small,
and then if such a state has finite energy
and positive entropy (e.g., by being a mixture
of the vacuum state $|0\!><\!0|$ and of $|\psi\!><\!\psi|$),
it can make $B \equiv S/(2\pi RE)$ arbitrarily large.

	If one did want to use $\varepsilon(\mathbf{x})$
as the weight function $W(\mathbf{x})$ in Eq. (\ref{eq:120}),
one would have to restrict the states so that $R_W^2$
cannot be too small for states of finite energy and
nonzero entropy.  One way that might work would be to
restrict the states to those in which $\varepsilon(\mathbf{x})$
is nonnegative everywhere, unlike the state $|\psi\!><\!\psi|$
for sufficiently large $-C\alpha/\beta$.

	Another option that might work for all
sufficiently localized states would be to choose
a weight function $W(\mathbf{x})$ that is nonnegative
for all states.  Examples of this would be
$\varepsilon^2$,
$\sum_{\mu=0}^4 \sum_{\nu=0}^4 (\tau_{\mu\nu})^2$,
$(\tau^{\mu}_{\mu})^2$,
$(\tau^{\mu}_{\nu}\tau^{\nu}_{\mu})^2$,
$(\tau^{\mu}_{\nu}\tau^{\nu}_{\rho}\tau^{\rho}_{\mu})^2$,
$(\tau^{\mu}_{\nu}\tau^{\nu}_{\rho}\tau^{\rho}_{\sigma}\tau^{\sigma}_{\mu})^2$,
\linebreak
$(tr(\phi(t\!=\!0,\mathbf{x})\rho))^2$,
$(tr(\dot{\phi}(t\!=\!0,\mathbf{x})\rho))^2$,
$(tr(:\!\phi^2(t\!=\!0,\mathbf{x})\!:\rho))^2$, 
$(tr(:\!\dot{\phi}^2(t\!=\!0,\mathbf{x})\!:\rho))^2$,
$(tr(:\!\phi^{;\mu}(t\!=\!0,\mathbf{x})
\phi_{;\mu}(t\!=\!0,\mathbf{x})\!:\rho))^2$,
etc., and positive powers of these positive functions
of $\mathbf{x}$ at $t=0$.

	The fourth quantity above is the square of
$\tau^{\mu}_{\nu}\tau^{\nu}_{\mu}$, which itself
usually seems to be positive everywhere,
though it might be some restriction of states
for this to be true everywhere for all states in the class.

	If one used one of these positive quantities,
or one of the usually positive quantities just for
states in which it is everywhere positive,
as a weight function $W(\mathbf{x})$ for defining
$R$ by Eq. (\ref{eq:120}),
one would again presumably get some upper bound
on the entropy $S$ as a function of $R$ and of $E$
(depending on how one defined $S$ and $E$
and what further restrictions one puts on the states).
However, I have not investigated what these
relations might be.

\section{Difficulties with Entropy Bounds
in Quantum and Semiclassical Gravity}

	So far I have been restricting attention
to nongravitational quantum field theories
in flat Minkowski spacetime.
However, since the original motivation
for Bekenstein's conjectured entropy bound
came from quantum considerations of
gravitating black holes,
it is interesting to consider whether
a similar entropy bound can be applied in quantum gravity.

	Here I must admit that I see serious problems
in attempting to apply the bound to quantum gravity.
Assuming that the quantum part of quantum gravity
is sufficiently similar to the ordinary quantum theory
of nongravitational systems, the entropy $S$
might still be a well-defined quantity,
at least for a complete system,
such as the entire universe
(though if the ultimate quantum gravity theory
specifies a unique quantum state,
there may be no option as to what the entropy is).
The energy $E$ is more problematic,
at least if the universe is not asymptotically flat,
though if one can restrict to quantum states in
which the universe is asymptotically flat,
then $E$ also might have a good definition
in quantum gravity.
However, what I don't see how to give a good
precise definition for is the size $R$.

	The main problem is that states of
quantum gravity should be coordinate invariant,
so it is hard to see how to say that some state
is confined to a radius $R$ or has this size.
How would one define the center with respect
to which the state is within a distance $R$?
Furthermore, if the state is an asymptotically
flat one with energy $E$, the gravitational field
of this energy should extend all the way out to
spatial infinity, so in that sense it seems
that the state cannot be confined to be within
radius $R$ or have vacuum properties outside
that radius.

	It is not that I have a rigorous proof
that an entropy bound such as Eq. (\ref{eq:1})
cannot be applied in quantum gravity,
but I just don't see how it can be applied.

	The situation seems somewhat more hopeful
in semiclassical gravity, in which one has quantum
field theory for nongravitational fields on a classical
curved spacetime (perhaps whose Einstein tensor
is proportional to the regularized expectation value
of the stress-energy tensor of the nongravitational
quantum fields).  In this case one can imagine
defining the equivalent of vacuum-outside-$R$
states of energy $E$ in the following way:

	Take an asymptotically flat
spherically symmetric spacetime
which has a totally geodesic Cauchy hypersurface
(with zero extrinsic curvature, say at $t=0$),
about which it has time-reflection symmetry ($t\rightarrow -t$).
Use a Schwarzschildean radial coordinate $r$,
the circumference/$(2\pi)$ of each symmetrical sphere.
Outside the $r=R$ sphere on this $t=0$ hypersurface,
assume that the spatial metric and the expectation
value of all operators confined to this region
are the same as that of the static spherically
symmetric asymptotically Schwarzschild
semiclassical metric with ADM mass $E$
and quantum state that is the semiclassical version
of the zero-temperature Boulware state
for this metric.  The entropy $S$ can then be
the von Neumann entropy
$S_{\rm vN} \equiv - tr\rho\ln\rho$
of the quantum state of the nongravitational
quantum field in the classical curved metric.

	If one applies this definition to
all possible states of this form,
then it seems one can easily violate
a bound of the form (\ref{eq:1})
by a state with arbitrarily large entropy
by having the $r=R$ sphere
be a neck separating the asymptotically flat
exterior with an interior on the $t=0$
hypersurface that is almost an entire
three-sphere of arbitrarily large size
filled with thermal radiation.
In other words, take the interior of
the $r=R$ two-sphere to be the moment
of maximum expansion of an almost-complete
large $k=1$ radiation Friedman-Robertson-Walker
model, and take the exterior to be
a moment of time-symmetry of a nearly empty
approximately Schwarzschild metric.
If the interior three-sphere radius is $a \gg R$,
giving an interior volume going as $a^3$,
the semiclassical Einstein equations imply that
the radiation energy density must go as $a^{-2}$
in Planck units at the moment of maximum expansion,
so the temperature $T$ goes as $a^{-1/2}$ and the entropy
density goes as $T^3$ or as $a^{-3/2}$.
When this is multiplied by the volume,
one gets an entropy going as $a^{3/2}$,
which can be made arbitrarily large
for fixed $R$ and $E$ (the asymptotic ADM mass)
by making the interior size $a$ arbitrarily large.

	One might seek to avoid this violation
of (\ref{eq:1}) by restricting the states to
which it is conjectured to apply to exclude
this example of a huge interior universe
separated from an asymptotically flat exterior
by a relatively small neck.

	One way to do that would be to demand that inside
the $r=R$ two-sphere on the hypersurface
of time symmetry, there are no round two-spheres
of radius greater than $R$
(i.e., topological two-spheres with intrinsic two-metrics
that are those of the standard unit round two-sphere
multiplied by a constant $r^2$ that is larger than $R^2$).
This would exclude an interior that is approximately
a large round three-sphere of radius $a \gg R$,
since such an interior region would have round two-spheres
of radii $r \approx a$.
However, it still appears to allow a very long throat
of radius near $R$, which by its arbitrarily great length
could have arbitrarily large volume and hence arbitrarily
large entropy $S$ for fixed $R$ and $E$.

	Another way to restrict the states
so that they might possibly obey a bound similar to (\ref{eq:1})
is to demand that the evolution of the semiclassical geometry
give a nonsingular metric over the whole of $\mathbf{R}^4$.
This would exclude the examples of a large internal
approximate three-sphere and also the long internal throat,
since these examples would be expected to collapse gravitationally
to singularities.  Only in cases in which the metric is not
too much different from flat spacetime would one expect that
no singularities develop from gravitational collapse,
and in these cases one might expect an entropy bound
not too different from its flat spacetime form.
However, for the restricted states to be sufficiently broad
to encompass most of the semiclassical gravity
generalizations of the allowed states (e.g., vacuum-outside-$R$
states) in the nongravitational theory,
the semiclassical Einstein equations should give
nonsingular evolution in these cases of sufficiently weak gravity.
Since the semiclassical Einstein equations are of higher
order in time than the ordinary Einstein equations with
a classical source, it is not clear that this will be the case,
and so one might need to find a suitable semiclassical gravity
theory before trying to apply entropy bounds.

\section{Conclusions and Acknowledgments}

	In conclusion, we have found that one can formulate precise
definitions for entropy bounds of a complete quantum field system (i.e.,
one not restricted to the interior of some boundary) by giving precise
definitions for the size $R$ of the system, at least when the metric is
classical so that sizes can be unambiguously defined. In particular, $R$
may be defined for vacuum-outside-$R$ states as the largest round
two-sphere on a suitable $t=0$ hypersurface, outside of which all of the
operators have the same expectation values as in a suitable vacuum state
(e.g., the ordinary vacuum state for nongravitational fields in
Minkowski spacetime, or a Boulware-type quantum state in semiclassical
gravity). Other values of $R$ may also be defined, such as the rms value
of $r$ with a suitable weight function dependent upon the quantum state
of the field. On the other hand, for a fully quantum gravity theory, it
appears to be difficult to give an unambiguous definition of a size $R$
of a system, so it is not clear there how to define a bound for the
entropy $S$ in terms of the energy $E$ and a size $R$.

	Discussions with Valeri Frolov, Jonathan Oppenheim, and L.
Sriramkumar have been helpful. This work was supported in part by the
Natural Sciences and Engineering Research Council of Canada.


\end{document}